\documentclass{article}
\pdfoutput=1
\usepackage{fullpage,hyperref}
\usepackage{xcolor}
\usepackage[T1]{fontenc}
\usepackage{lmodern}
\hypersetup{
    colorlinks,
    linkcolor={red!50!black},
    citecolor={blue!50!black},
    urlcolor={blue!80!black},
    pdfauthor={José A. Soto},
    pdftitle={Improved Analysis of a Max-Cut Algorithm Based on Spectral Partitioning.}
}
\usepackage{mathtools,amsmath}
\usepackage{amsthm,amssymb}
\usepackage{algpseudocode,algorithm}

\algdef{SE}[IF]{NoThenIf}{EndIf}[1]{\algorithmicif\ #1}{\algorithmicend\ \algorithmicif}%

\newtheorem{theorem}{Theorem}[section]
\newtheorem{lemma}[theorem]{Lemma}

\theoremstyle{definition}

\theoremstyle{remark}
\newtheorem{rem}[theorem]{Remark}
\numberwithin{equation}{section}

\newcommand{\R}{\mathbb R}

\newcommand{\Adj}{\mathrm{Adj}}
\newcommand{\Deg}{\mathrm{Deg}}
\newcommand{\A}{\mathcal{A}}
\newcommand{\E}{\mathbb{E}}
\newcommand{\Good}{\mathrm{Good}}
\newcommand{\Bad}{\mathrm{Bad}}
\newcommand{\Cross}{\mathrm{Cross}}
\newcommand{\Inc}{\mathrm{Inc}}
\begin{document}
\title{Improved Analysis of a Max-Cut Algorithm Based on Spectral Partitioning.}

\author{Jos\'e A.~Soto\thanks{
Departamento de Ingenier\'ia Matem\'atica and Centro de Modelamiento Matem\'atico (UMI
2807 CNRS), Universidad de Chile. \url{jsoto@dim.uchile.cl}. Supported by FONDECYT 11130266 and N\'ucleo Milenio Informaci\'on y Coordinaci\'on en Redes ICM/FIC P10-024F.}}

\maketitle
\begin{abstract}
Trevisan [SICOMP 2012] presented an algorithm for Max-Cut based on spectral partitioning techniques. This is the first algorithm for Max-Cut with an approximation 
guarantee strictly larger than 1/2 that is not based on semidefinite programming. Trevisan showed that its approximation ratio is of at least 0.531. In  this paper we improve this bound up to 0.614247. We also define and extend this result for the more general Maximum Colored Cut problem.
\end{abstract}

\section{Introduction}
The maximum cut (Max-Cut) problem consists of finding a bipartition of the vertices of a weighted graph that maximizes the total weight of the edges crossing it. Max-Cut is one of Karp's original NP-complete problems, so finding heuristics and approximation algorithms for it has 
attracted researchers for years. It is easy to find a solution for Max-Cut whose weight is at least half of the optimum: a partition chosen 
uniformly at random will cut, in expectation, half of the total weight of the graph, and this process can be derandomized using standard 
techniques.
No approximation asymptotically better than 1/2 was known for this problem until Goemans and Williamson \cite{GW} devised an algorithm based on 
semidefinite programming (SDP) yielding a $0.87856$ approximation. Although  SDP is 
solvable in polynomial time, it is still computationally expensive in practice, so it is an intriguing question to develop approximation algorithms 
for Max-Cut that do not use SDP. Semidefinite programming was the only known method to achieve a guarantee of more than a half until 
Trevisan~\cite{Luca} devised an algorithm based on eigenvector computations whose approximation factor is at least 0.531. In this article, we present 
a new analysis of Trevisan's algorithm that improves that bound up to 0.614247. We also apply a modification of his algorithm to the slightly more 
general maximum colored cut problem.

\paragraph{Organization} In Section 2 we describe the maximum colored cut problem and our notation. We also prove our main technical lemma, 
Lemma~\ref{lemaimportante} which is the heart of the new analysis. In Section 3 we revisit 
Trevisan's algorithm and prove our new guarantee.

\section{Maximum Colored Cut Problem}
In the \emph{maximum colored cut} problem (MaxCC), we are given a graph $G=(V,E)$ with $E=R\cup B$, $R\cap B=\emptyset$.
The edges in $R$ are said to be  red and the edges in $B$ are said to be blue. We are also given a nonnegative weight function $w\colon E \to \R_{\geq 0}$. The goal is to partition $V$ into two sets $V_-$ and $V_+$ such that the weight of the red edges that are cut by the partition plus the weight of the blue edges that are not cut (or \emph{uncut}) is maximized. This problem generalizes the Max-Cut problem, in which $B=\emptyset$.

Given a bipartition of $V$, we say that a red edge is \emph{good} if it belongs to the associated cut, and that a blue edge is \emph{good} if it is 
not in the cut. Any edge which is not good is called \emph{bad}. With this notation, the objective of the MaxCC problem is to find a partition 
that maximizes the total weight of good edges.

Trevisan's approach for Max-Cut~\cite{Luca} can be described as follows: Devise an algorithm $\A$ that finds a tripartition $(V_-,V_0,V_+)$ of $V$, 
where $V_0\neq V$ is the set of nodes that are still ``undecided'' between $V_-$ and $V_+$ and then recursively use $\A$ on the set of undecided nodes until 
every node is decided. Observe that for a given tripartition $(V_-,V_0,V_+)$, the edges of the graph induced by $V_- \cup V_+$ are already labeled as 
good or bad, and no matter what the recursive partition of the undecided vertices into two pieces $W_-$ and $W_+$ is, we can always impose that half of 
the total weight of the edges between $V_-\cup V_+$ and $V_0=W_- \cup W_+$ is good, by assigning $W_-$ to $V_-$ and $W_+$ to $V_+$ or vice versa. This 
observation suggests that the objective of $\A$ should be to find a tripartition for which the ratio of the weight of the good edges induced by $V_- 
\cup V_+$ plus half of the total weight crossing from $V_- \cup V_+$ to $V_0$ with respect to the total weight of the edges involved (the ones 
incident to $V_- \cup V_+$) is as high as possible. We call this ratio the \emph{recoverable ratio} of the tripartition. As we will later see, it is 
possible to find a partition with high recoverable ratio by using spectral partitioning techniques.

The algorithm we describe and analyze in Section 3 is the same as the one by Trevisan in~\cite{Luca}, so we assume certain familiarity with that 
paper. The previous analysis for the approximation ratio of that algorithm involve upper bounding the number of uncut edges (bad edges, in our 
nomenclature) in each iteration via an application of Cauchy-Schwarz inequality. We give a tighter analysis by directly lower bounding the 
number of cut edges (good edges).

\subsection{Notation}
For a graph $G=(V,E)$, a set of edges $F \subseteq E$, and a set of vertices $A \subseteq V$, we use $F[A]$ to denote the set of edges in $F$ with both endpoints inside of $A$. Similarly, for disjoint subsets $A_1$ and $A_2$ of $V$ we use $F(A_1:A_2)$ to denote the set of edges in $F$ with one endpoint in $A_1$ and the other in $A_2$. The indicator vector of a bipartition $(V_-,V_+)$ of $V$ is the vector $x \in \{-1,1\}^V$ with $x_i = -1$ if $i \in V_-$ and $x_i=1$ if $i \in V_+$. Similarly, the indicator vector of a tripartition $(V_-,V_0,V_+)$ of~$V$ is the vector $z \in \{-1,0,1\}^V$ with $z_i$ equals to $-1$, $0$ or $1$ whenever $i$ is in $V_-$, $V_0$ or~$V_+$, respectively. As usual, for a weight function $w\colon E \to \R_{\geq 0}$, and a set $F \subseteq E$, we use $w(F)$ as a shorthand for $\sum_{e \in F}w(e)$. For the rest of the paper, we fix a graph $G=(V,E)$ with $V=[n]=\{1,\dots,n\}$, $E=R\cup B$, $R\cap B = \emptyset$, and a nonnegative weight function $w\colon E \to \R_{\geq 0}$.

The following quadratic formulation gives the value of the maximum colored cut.
\begin{align}
  \max_{x \in \{-1,1\}^n} \frac{1}{4} \sum_{\{i,j\} \in R} w_{ij} (x_i - x_j)^2 + \frac{1}{4}\sum_{\{i,j\} \in B} w_{ij} (x_i + x_j)^2. \label{MCC} \tag{MaxCC}
\end{align}

Let $M^{\{i,j\}} \in \R^{n \times n}$ be the matrix associated to the quadratic form involving edge $\{i,j\}$ for MaxCC:
\begin{align*}
 x^T M^{\{i,j\}} x= \begin{cases}
    w_{ij} (x_i - x_j)^2& \text{ if $\{i,j\} \in R$},\\
    w_{ij} (x_i + x_j)^2& \text{ if $\{i,j\} \in B$.}
  \end{cases}
\end{align*}
By letting $M=\sum_{\{i,j\}\in E} M_{ij}$, MaxCC can be expressed as $\max_{x \in \{-1,1\}^n} \frac{1}{4}x^T M x$. 

It will also be convenient to define matrices $D^{\{i,j\}}$ such that
\begin{align*}
 x^T D^{\{i,j\}} x&=
    w_{ij} (x_i^2 + x_j^2),
\end{align*}
and let $D = \sum_{\{i,j\} \in E} D^{\{i,j\}}$. Note that for every $x \in \{-1,1\}^n$,  $x^TDx = 2w(E)$.
It is also easy to check that
\begin{align*}
M &= \Deg(R) - \Adj(R) + \Deg(B) + \Adj(B),\\
D &= \Deg(R) + \Deg(B),
\end{align*}
where $\Adj(K)$ and $\Deg(K)$ are the weighted adjacency and degree matrices associated to the edge set $K$.

The following lemma, which is a restatement of Lemma 2 of~\cite{Luca} in our terminology, relates the value of the MaxCC to the highest eigenvalue of a certain positive semidefinite matrix.

\begin{lemma}\label{eigen}
If $\text{MaxCC} \geq (1-\varepsilon)w(E)$, then there exists a vector $x \in \R^n\setminus\{0\}^n$ such that
\begin{align*}
\frac{x^T M x}{x^T D x} \geq 2(1-\varepsilon).
\end{align*}
\end{lemma}
\begin{proof}
Let $x$ be the vector solving  $\max_{x\in \R^n} x^T M x/x^T D x$, and $x^*\in \{-1,1\}^n$ be the incident vector of a maximum colored cut. Then
$$\frac{x^T M x}{x^T D x} \geq \frac{{x^*}^T M x^*}{{x^*}^T D x^*} = \frac{4\text{MaxCC}}{2 w(E)} \geq 2(1-\varepsilon).$$
To find $x$ we can simply find a unit eigenvector $y$ associated to the maximum eigenvalue of the positive semidefinite matrix $D^{-1/2}MD^{-1/2}$, and then set $x=D^{-1/2}y$. \qquad
\end{proof}

\medskip
In practice, the vector $x$ guaranteed by the previous lemma can not be found exactly in polynomial time. However, it is possible to efficiently find a vector $x$ such that $x^T M x\geq 2(1-\varepsilon - 
\delta)x^T D x$ in time inversely proportional to $\delta$ (see the discussion of Lemma 2 in \cite{Luca}). 
We show next how to find a good tripartition, this is, one having a good recoverable ratio, using vector $x$ as a starting 
point.

\subsection{Finding a good tripartition}

Let $z \in \{-1,0,1\}^n\setminus \{0\}^n$ be the indicator vector of a partition $(V_-,V_0,V_+)$ with $V\neq V_0$. Define the weights of the good, bad, crossing  and incident edges associated to the partition as:
\begin{align*}
\Good(z) &:= w(R(V_-:V_+)) + w(B[V_-] \cup B[V_+]),\\
\Bad(z)  &:= w(B(V_-:V_+)) + w(R[V_-] \cup R[V_+]),\\
\Cross(z)&:= w(E(V_- \cup V_+:V_0)),\\
\Inc(z)  &:= w(E)-w(E[V_0]),
\end{align*}
respectively. The next two observations are direct from the previous definitions:
\begin{align}
\Inc(z)&=\Good(z) + \Bad(z) + \Cross(z),\label{eqn:21}\\
z^TDz&=\sum_{\{i,j\} \in E}w_{ij}(z_i^2+z_j^2)=\sum_{e \in E[V_-\cup V_+]} 2w_{ij}+\sum_{e\in E(V_- \cup V_+:V_0)}w_{ij}\label{eqn:22} \\
&=2\Inc(z)-\Cross(z).\notag
\end{align}

The \emph{recoverable ratio} of $z$ is defined to be  $(\Good(z) + \Cross(z)/2)$ 
divided by $\Inc(z)$, if the denominator is not zero. If $\Inc(z)=0$, this ratio is defined\footnote{The definitions of this section only apply to nonzero vectors $z$. Even if $z\neq 0$, $\Inc(z)$ may be zero, but for that to happen we need that all the edges incident to $V\setminus V_0 \neq \emptyset$ have zero weight. In this case, the vertices in $V\setminus V_0$ are irrelevant for the maximium colored cut instance.  Note that this always occurs in certain situations, for instance if $|V|=1$.} as 1. Note that if $x^*$ is the indicator vector of an optimal colored cut, (in particular, $V_0=\emptyset$), its recoverable ratio is exactly 
$\text{MaxCC}/w(E) = {x^*}^TMx^*/(2{x^*}^TD{x^*})$. 

Given a vector $x \in \R^n\setminus\{0\}^n$ such that $x^TMx/ x^TDx$ is large, we want to find a rounded vector $z \in \{-1,0,1\}^n \setminus \{0\}^n$ with high recoverable ratio (the condition that $z$ is nonzero is added explicitly in order to avoid situations in which all vertices are left ``undecided'').
The following lemma, which can be seen as an improvement of Lemma 3 in \cite{Luca}, is useful for this task.
\begin{lemma}\label{rounding}
Let $x \in \R^n$, with $\|x\|_\infty = 1$. Construct $y \in \{-1,0,1\}^n\setminus\{0\}^n$ as follows. Pick $t$ uniformly at random from the open interval $(0,1)$ and let, for each $i$,
  $$y_i = \begin{cases}
    1,& \text{if $x_i \geq \sqrt{t}$}.\\
    -1,& \text{if $x_i \leq -\sqrt{t}$}.\\
    0,&  \text{if $|x_i| \leq \sqrt{t}$}.
  \end{cases}$$
Let \begin{align*}
C(i,j) &=\Pr\left[\binom{y_i}{y_j} \in \left\{\binom{-1}{1},\binom{1}{-1}\right\}\right],\\
U(i,j) &=\Pr\left[\binom{y_i}{y_j} \in \left\{\binom{1}{1},\binom{-1}{-1}\right\}\right], \text{and }\\
X(i,j) &=\Pr\left[\binom{y_i}{y_j} \in \left\{\binom{0}{1},\binom{0}{-1},\binom{1}{0},\binom{-1}{0}\right\}\right],
\end{align*}
be the probabilities that an edge $\{i,j\}$ is cut, uncut or crossing the tripartition induced by $y$, respectively.
Then for all $0\leq \beta \leq 1$,
\begin{align}
U(i,j) + \beta X(i,j) &\geq \beta(1-\beta) (x_i + x_j)^2, \label{eq1}\\
C(i,j) + \beta X(i,j) &\geq \beta(1-\beta) (x_i - x_j)^2.\label{eq2}\end{align}
\end{lemma}

\begin{proof}
First note that the random vertex $y$ is nonzero since for the indices $i$ such that $|x_i|=1$, we also have $|y_i|=1$. Also observe that \eqref{eq2} follows if we apply \eqref{eq1} to the vector $x'$ obtained from $x$  by switching the sign of $x_j$.
To prove \eqref{eq1} we consider two cases.

\smallskip
\noindent \textbf{Case 1}: $x_i x_j \geq 0$. Assume, w.l.o.g.~that $|x_i| \leq |x_j|$. In this case, $U(i,j)$ equals the probability that 
both $x_i^2$ and $x_j^2$ are bigger than $t$, thus $U(i,j) = x_i^2$. Similarly, $X(i,j)$ is equal to the probability that $t$ is between $x_i^2$ and 
$x_j^2$. This is, $X(i,j) = (x_j^2 -x_i^2)$.

Using a version of  Bergstr\"om's inequality (see, e.g.~\cite{inequalities}), 
$$\beta(1-\beta)(a+b)^2 \leq 
(1-\beta) a^2 + \beta b^2,$$ which is valid for $a,b \geq 0$ and $0\leq \beta \leq 1$, we obtain that
\begin{align*}
  \beta(1-\beta)(x_i + x_j)^2 =   \beta(1-\beta)(|x_i| + |x_j|)^2 \leq (1-\beta)x_i^2 + \beta x_j^2 = U(i,j) + \beta X(i,j).
\end{align*}

\noindent \textbf{Case 2}: $x_i x_j < 0$. Assume again, w.l.o.g.~that $|x_i| \leq |x_j|$. It is easy to see that $U(i,j) = 0$ 
and that $X(i,j) = (x_j^2 -x_i^2)$.

Since $x_i x_j < 0$, we have $|x_j + x_i| \leq |x_j - x_i|$. Using that $0\leq \beta \leq 1$, we get
\begin{equation*}
  \beta(1-\beta)(x_i + x_j)^2 \leq \beta|x_i + x_j|\cdot |x_i - x_j| = \beta(x_j^2-x_i^2) = U(i,j) + \beta X(i,j).  \qedhere
\end{equation*}
\end{proof}

\bigskip

The main technical result of this article is given by the following lemma.

\begin{lemma}\label{lemaimportante}
Given a nonzero vector $x$ such that  $x^T M x \geq 2(1-\varepsilon)\cdot x^TDx$, for some $0 \leq \varepsilon \leq 1/2$, we can efficiently find  
an indicator vector $z \in \{-1,0,1\}^n\setminus \{0\}^n$ of a tripartition satisfying
\begin{align} \label{desired}
\Good(z) + \frac{\Cross(z)}{2} \geq \begin{cases}
  \displaystyle\frac{-1 +\sqrt{4\varepsilon^2 -8\varepsilon + 5}}{2(1-\varepsilon)}\cdot \Inc(z) &\text{if $\varepsilon\geq \varepsilon_0$,}  \\[10pt] \displaystyle\frac{1}{1+2\sqrt{\varepsilon(1-\varepsilon)}}\cdot \Inc(z) &\text{if $\varepsilon\leq \varepsilon_0$.}
\end{cases}
\end{align}
where $\varepsilon_0\approx 0.22815\ldots $ is the unique solution of the equation $$\displaystyle \frac{1}{1+2\sqrt{\varepsilon(1-\varepsilon)}} = 
\frac{-1 +\sqrt{4\varepsilon^2 -8\varepsilon + 5}}{2(1-\varepsilon)}.$$

\end{lemma}

\begin{proof}
Without loss of  generality, assume that $\|x\|_\infty = 1$. Let $y$ be the random vector obtained from $x$ as in Lemma \ref{rounding}. Define the expected recoverability of $x$ as $\rho(x)=\E\left[\Good(y) + \Cross(y)/2\right]\big/\,\E[\Inc(y)]$ if $\E[\Inc(y)]\neq \emptyset$, and $\rho(x)=1$ otherwise. In what follows we find some lower bounds for $\rho(x)$.

Let $0\leq \beta \leq 1$ to be specified later. Using Lemma \ref{rounding}, the identity $x^TD x =\sum_{\{i,j\} \in E}w_{ij}(x_i^2+x_j^2) =\allowbreak \sum_{\{i,j\} \in E}w_{ij}\E[y_i^2+y_j^2]=\E[y^TDy]$, and equation \eqref{eqn:22}  we obtain that $\E[\Good(y) + \beta \Cross(y)]$ equals
 \begin{align*}
\sum_{\{i,j\} \in R}& w_{i,j}(C(i,j) + \beta X(i,j)) + \sum_{\{i,j\} \in B} w_{i,j}(U(i,j) + \beta X(i,j))\\
&\geq \beta(1-\beta) \sum_{\{i,j\} \in R} w_{i,j}(x_i - x_j)^2 + \beta(1-\beta) \sum_{\{i,j\} \in B} w_{i,j}(x_i + x_j)^2\\
&= \beta(1-\beta) x^T M x \geq 2(1-\varepsilon)\beta(1-\beta)  x^T D x.\\
&= 2(1-\varepsilon)\beta(1-\beta)\E[2\Inc(y)-\Cross(y)].
\end{align*}
Rearranging terms we get 
\begin{align} \label{eqbeta}
2A\E[\Inc(y)] &\leq \E[\Good(y)] + (A +\beta)\E[\Cross(y)],
\end{align}
where $A=2(1-\varepsilon)\beta(1-\beta)$. By \eqref{eqn:21} we obtain that 
$\Cross(y)=\Inc(y)-\Good(y)-\Bad(y)\leq \Inc(y)-\Good(y)$; therefore, for every $\alpha\geq 0$,
\begin{align} 
(2A-\alpha)\E[\Inc(y)] &\leq \E[\Good(y)](1-\alpha) + (A+\beta-\alpha)\E[\Cross(y)].
\end{align}

Imposing $1-\alpha = 2(A+\beta-\alpha)$, which requires that $\alpha=2(A+\beta) -1\geq 0$,  we get
\begin{align} 
(1-2\beta)\E[\Inc(y)] &\leq 2(1-A-\beta)\left(\E[\Good(y)]+ \frac12 \E[\Cross(y)]\right).
\end{align}
 Finally, further imposing that $A+\beta<1$, we obtain
\begin{align} 
\frac{1-2\beta}{2(1-A-\beta)}\E[\Inc(y)] &\leq \E[\Good(y)]+ \frac12 \E[\Cross(y)].
\end{align}

It follows that $\rho(x)$ is at least the maximum of $(1-2\beta)/(2(1-A-\beta))$ subject to $1/2\leq A+\beta<1 $, and $0\leq \beta \leq 1$. By  the definition of $A$ and using that $0\leq \varepsilon \leq 1$, we get that $\rho(x)$ is at least the maximum of
\begin{align*}
R(\beta,\varepsilon)&:=\frac{1-2\beta}{2(1-\beta)(1-2\beta(1-\varepsilon))},\\ \text{
subject to }\quad
\beta_{\min}(\varepsilon)&:=\frac{1}{3-2\varepsilon+\sqrt{5-8\varepsilon +4\varepsilon^2}} \leq  \beta < \frac{1}{2(1-\varepsilon)}.
 \end{align*}

For fixed $\varepsilon$, the function $R(\beta,\varepsilon)$ is continuous and differentiable in $\beta$; therefore $R(\cdot,\varepsilon)$  achieves its maximum either in a point  with $\partial  R(\beta,\varepsilon)/ \partial \beta =0$ or in the point $\beta_{\min}(\varepsilon)$ that is in the border of the region defined by the constraints.  Note that $\partial  R(\beta,\varepsilon)/ \partial \beta =0$ if and only if $\beta\in\{\beta_-(\varepsilon),\beta_+(\varepsilon)\}$, where $\beta_\pm(\varepsilon):=\frac{1}{2}\pm\frac{1}{2}\sqrt{\varepsilon/(1-\varepsilon)}$. The point $\beta_+(\varepsilon)$ is always outside the feasible region, while the point $\beta_-(\varepsilon)$ belongs to $[\beta_{\min}(\varepsilon), 1/(2(1-\varepsilon)))$ if and only if  $\varepsilon \geq \varepsilon_0 \approx 0.22815$, where $\varepsilon_0$ is the unique value that makes $\beta_-(\varepsilon)=\beta_{\min}(\varepsilon)$.

Consider now the functions
\begin{align}
f_1(\varepsilon) = R(\beta_{\min}(\varepsilon))&=\frac{-1+\sqrt{4\varepsilon^2-8\varepsilon+5}}{2(1-\varepsilon)},\qquad  f_2(\varepsilon) = R(\beta_-(\varepsilon)) =\frac{1}{1+2\sqrt{\varepsilon(1-\varepsilon)}},\notag \\[5pt]
\text{and } f(\varepsilon) &= \begin{cases}
  f_1(\varepsilon) , &\text{ if $\varepsilon \geq \varepsilon_0$,}\\
  f_2(\varepsilon), &\text{ if $\varepsilon \leq \varepsilon_0$.}\end{cases}\label{function-f}
\end{align}
From the previous discussion, we deduce that if $\E[\Inc(y)]\neq 0$, $\rho(x)\geq f(\varepsilon)\geq 0$. Therefore,
$$\E\left[\Good(y) + \frac{\Cross(y)}{2} - f(\varepsilon) \E[\Inc(y)]\right]\geq 0.$$

Observe that the previous inequality holds even if $\E[\Inc(y)]=0$. Since the expectation above is always nonnegative, there must exist a vector $z \in \{-1,0,1\}^n\setminus\{0\}^n$, among those that participate in the expectation, with recoverable ratio at least $f(\varepsilon)$. We can find $z$ by testing all $n$ thresholds $\{x_i^2\}_{i=1}^n$ for $t$ in the proof of Lemma~\ref{rounding}, and keeping the associated vector with greatest recoverable ratio. Note that the vector $z$ obtained like this is not zero, since for the index $i$ such that $|x_i| = 1$, we must also have $|z_i| = 1$. \qquad
\end{proof}

\begin{figure}[ht!]
\begin{minipage}[t]{.5\textwidth}
  \centering
  \includegraphics[scale=0.5]{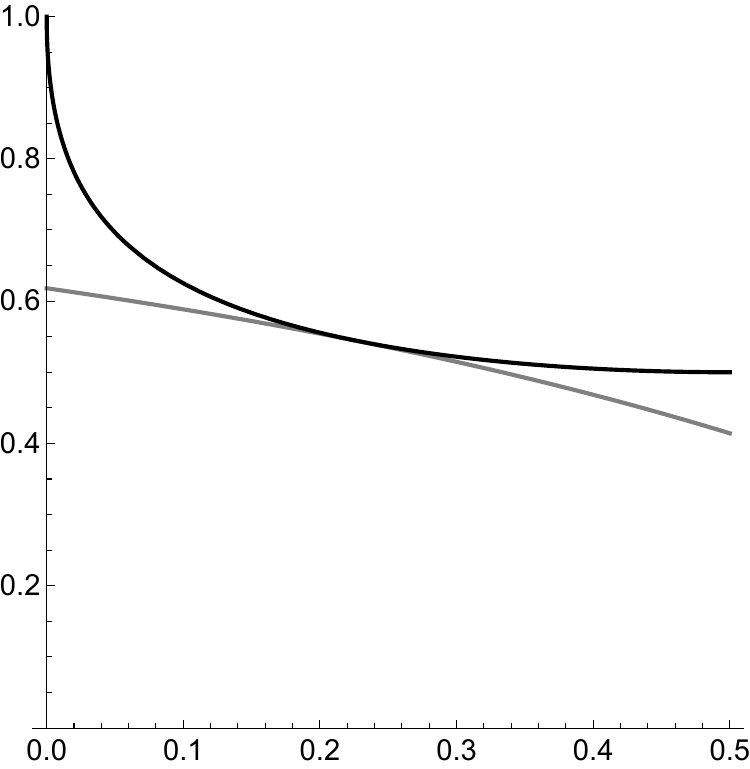}
\end{minipage}%
\begin{minipage}[t]{.5\textwidth}
  \centering
  \includegraphics[scale=0.5]{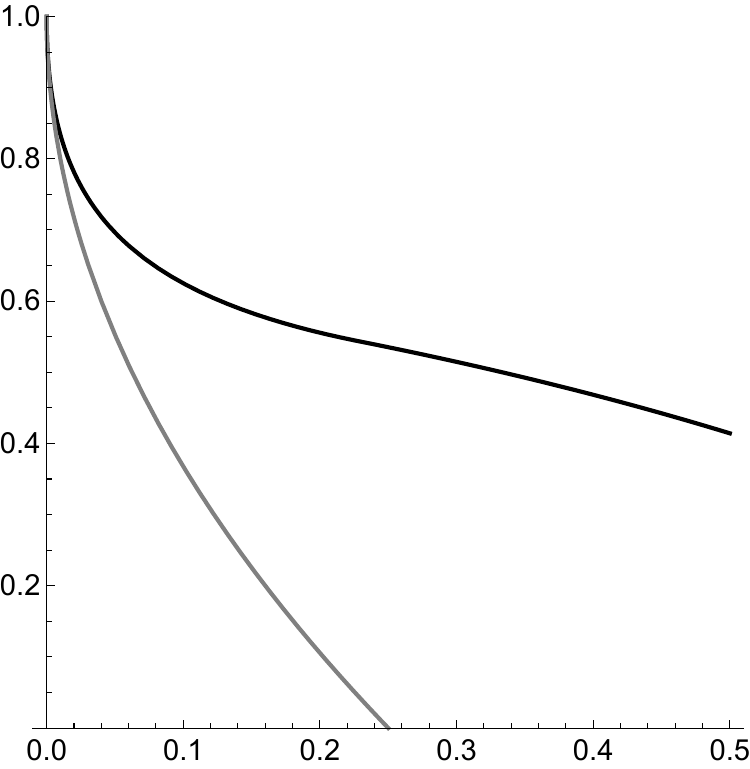}
\end{minipage}
  \caption{On the left, the functions $f_1(\varepsilon)$ (bottom function) and $f_2(\varepsilon)$ (top function) used to defined $f(\varepsilon)$. On the right, the relationship between the recoverable ratio guaranteed by $f(\varepsilon)$ (top function), and that guaranteed by $\tilde{f}(\varepsilon)$ (bottom function). }
  \label{fig:plot1}
\end{figure}
\medskip
\begin{rem}\label{remark}
{\rm Lemmas \ref{eigen} and \ref{lemaimportante} imply that if $G$ has a bipartition with at least $1-\varepsilon$ fraction of the total weight being good (`cut'), then we can  find a partition into three pieces $V_-,V_0,V_+$ with recoverable ratio at least $f(\varepsilon)$. 
Previously, Trevisan proved, using an application of Cauchy-Schwarz inequality, that the same is true for the function $\tilde{f}(\varepsilon) = 1-2\sqrt{\varepsilon}$. In Figure~\ref{fig:plot1} we can see the relationship between these two bounds. Note that we always have $f(\varepsilon) \geq \tilde{f}(\varepsilon)$. Also, it is worth noting that a random bipartition has a recoverable ratio of 1/2 in expectation. The previous guarantee $\tilde{f}(\varepsilon)$ beats 1/2 when $\varepsilon<1/16$, while the new guarantee $f(\varepsilon)$ implies that we can do better than a random cut for any $\varepsilon < 1/3$.}

\end{rem}

\section{Trevisan's algorithm and our new analysis}
Our extension of Trevisan's algorithm for the MaxCC problem is depicted as Algorithm~\ref{alg:trevisan} below. In what follows, we assume that the vector $x$ given by Lemma \ref{eigen} can be found exactly in order to keep the argument simpler.

\begin{algorithm}[!ht]
\caption{(Spectral Partitioning).}
\label{alg:trevisan}
\begin{algorithmic}[1]
\Require{A colored graph $(V,E=R\cup B)$, with nonnegative weights $w\colon E\to\R_{\geq 0}$ .}
\Ensure{A bipartition $\texttt{ALG}(V,R,B,w):=(V_-,V_+)$ of $V$.}
\State{Compute a nonzero vector $x$ maximizing $x^TMx / x^TDx$ (Lemma \ref{eigen}).}
\State{Determine from $x$ the rounded vector $z\in \{-1,0,1\}^{V}\setminus\{0\}^{V}$ given by Lemma \ref{lemaimportante}}
\If{$\displaystyle \Good(z) + \Cross(z)/2< \Inc(z)/2$}\State{Return a bipartition $(V_-,V_+)$ of $V$ such that the weight of good edges is at least half of $w(E)$\, (a random cut suffices).}
\Else\State{Let $(V_-,V_0,V_+)$ be the tripartition induced by $z$.}
\If{$V_0 = \emptyset$} 
\State{Return $(V_-,V_+)$.}
\Else 
\State{Set $(W_-,W_+) \gets \texttt{ALG}(V_0,R[V_0],B[V_0],w|_{E[V_0]})$.}
\State{Return the best of
  $(V_- \cup W_-,V_+\cup W_+)$ and $(V_- \cup W_+,V_+ \cup W_-)$.} \EndIf\EndIf
\end{algorithmic}
\end{algorithm}

\begin{theorem}\label{theoremfinal}
If $\textrm{MaxCC}(G) = (1- \varepsilon) w(E)>0$, then Algorithm~\ref{alg:trevisan} returns a partition $(V_-,V_+)$ whose indicator vector $y \in \{-1,1\}^n$ satisfies
$$\frac{\Good(y)}{w(E)} \geq \int_0^1\max\left(\frac12,f(\varepsilon/r)\right)dr$$
with $f(\varepsilon)$ as in \eqref{function-f}.
\end{theorem}

\begin{proof} 
Note first that the algorithm terminates since in every recursive call the residual graph $(V_0,E[V_0])$ considered has at least one fewer node. Assume that the algorithm performs $T$ recursive calls and let $G_t=(V_t,E_t)$ be the graph at the beginning of the $t$-th iteration, so that $G_1=(V,E)$. Let also $G_{T+1}=(\emptyset,\emptyset)$ be the empty graph and $\delta_t = w(E_t)/w(E)$ for every $t$. We observe that
  $$\frac{w(E_t)-\text{MaxCC}(G_t)}{w(E_t)} \leq \frac{w(E)-\text{MaxCC}(G)}{\delta_t \cdot w(E)}\leq \frac{\varepsilon}{\delta_t}.$$
The previous holds since for the optimal bipartition of $V$ defining $\text{MaxCC}(G)$, the total weight of the bad edges is at least the weight of the bad edges inside $E_t$ for the same partition, which in turn  is at least the weight of the bad edges for the bipartition of $V_t$ defining $\text{MaxCC}(G_t)$. From the previous observation we get that  $\text{MaxCC}(G_t) \geq (1-\varepsilon/\delta_t) w(E_t)$. 

Recall that a random cut has a recoverable ratio of at least half.  Combining this with Remark~\ref{remark} and the intuition about the recoverable ratio given in the introduction we get that the total weight of good edges in $E_t \setminus E_{t+1}$ with respect to the partition $(V_-,V_+)$ is at least 
 $$\max\left(\frac{1}{2},f(\varepsilon/\delta_t)\right)\cdot w(E_t\setminus E_{t+1}).$$
Using that $f$ is decreasing, and that the sets $E_t\setminus E_{t+1}$ are mutually disjoint, we obtain that the total good weight returned by the algorithm is at least
\begin{align*}
\frac{\Good(y)}{w(E)} &\geq \sum_{t=0}^{T}\max\left(\frac{1}{2},f(\varepsilon/\delta_t)\right)\cdot (\delta_t - \delta_{t+1})\\
&= \sum_{t=0}^{T} \int_{\delta_{t+1}}^{\delta_t}\max\left(\frac{1}{2},f(\varepsilon/\delta_t)\right)dr\\
&\geq \sum_{t=0}^{T} \int_{\delta_{t+1}}^{\delta_t}\max\left(\frac{1}{2},f(\varepsilon/r)\right)dr\\
&= \int_{0}^{1}\max\left(\frac{1}{2},f(\varepsilon/r)\right)dr. \qedhere
\end{align*}
\end{proof}

\begin{theorem}
Algorithm~\ref{alg:trevisan} guarantees a 0.6142 approximation for MaxCC.
\end{theorem}

\begin{proof}
Let $F(\varepsilon)=\int_{0}^{1}\max\left(1/2,f(\varepsilon/r)\right)dr$. By Theorem \ref{theoremfinal}, $F(\varepsilon)w(E)$ is a lower bound on the good weight returned by Algorithm~\ref{alg:trevisan}. Note that $f(x) = 1/2$ when $x=1/3$. Using this, and the definition of $f$ it is easy to see that

\begin{align*}
F(\varepsilon)&= \begin{cases}
  \displaystyle \frac12, &\text{ if $\varepsilon \geq 1/3$}.\vspace{10pt}\\
\displaystyle \int_0^{3\varepsilon} \frac12 dr + \int_{3\varepsilon}^1 \frac{-1+\sqrt{4(\varepsilon/r)^2-8(\varepsilon/r)+5}}{2(1-\varepsilon/r)}dr, &\text{ if $\varepsilon_0 \leq \varepsilon \leq 1/3$.}\vspace{10pt}\\
\displaystyle \int_0^{3\varepsilon} \frac12 dr + \int_{3\varepsilon}^{\varepsilon/\varepsilon_0} \frac{-1+\sqrt{4(\varepsilon/r)^2-8(\varepsilon/r)+5}}{2(1-\varepsilon/r)}dr \\[10pt]
\qquad + \displaystyle \int_{\varepsilon/\varepsilon_0}^1\frac{1}{1+2\sqrt{(\varepsilon/r)(1-\varepsilon/r)}}dr , &\text{ if $\varepsilon \leq \varepsilon_0$.\vspace{10pt}}
\end{cases}
\end{align*}
with $\varepsilon_0 \approx 0.228155$, as in Lemma \ref{lemaimportante}.

The approximation guarantee, as a function of $\varepsilon$  is  given by $G(\varepsilon)=F(\varepsilon)/(1-\varepsilon)$. We can check that the function $G(\varepsilon)$ is convex and has an unique minimum at $\varepsilon^* \approx .11089$ with value $G(\varepsilon^*) \approx 0.614247$. \qquad
\end{proof}

\medskip
For completeness, we include in the next section a closed form of function $G(\varepsilon)$ and a plot comparing this guarantee with the previous guarantee of Trevisan~\cite{Luca} for different regimes of $\varepsilon$.

\subsection{Analytic expression of the guarantee function}

If $\text{MaxCC}(G)\geq w(E)\cdot (1-\varepsilon)$, then the algorithm described above gives a $G(\varepsilon)$-approximation for the MaxCC problem, with $G(\varepsilon)$ defined by the following expression:

\noindent If $1/3\leq \varepsilon\leq 1/2$,
\begin{align*}
  G(\varepsilon):= \frac{1}{2(1-\varepsilon)}.
\end{align*}
If $\varepsilon_0\leq \varepsilon \leq 1/3$,
\begin{align*}
 G(\varepsilon):=\frac{1}{2(1-\varepsilon)}\cdot \bigg(&\varepsilon-1  + \sqrt{4\varepsilon^2-8\varepsilon+5}-\varepsilon\ln\left(\frac{1+\sqrt{4\varepsilon^2-8\varepsilon+5}}{8\varepsilon}\right)\\
 &+\frac{\sqrt{5}}{5}\varepsilon\ln\left(\frac{5-4\varepsilon+\sqrt{5(4\varepsilon^2-8\varepsilon+5)}}{\left(11+5\sqrt{5}\right)\varepsilon}\right)\bigg);
\end{align*}
and if $0\leq \varepsilon \leq \varepsilon_0$,
\begin{align*}
G(\varepsilon):=\frac{1}{2(1-\varepsilon)}&\cdot \Bigg(\varepsilon\left(1-\frac{3}{\varepsilon_0}\right)+2+\frac{\varepsilon}{\varepsilon_0}\sqrt{4\varepsilon_0^2-8\varepsilon_0+5}-\varepsilon\ln\left(\frac{1+\sqrt{4\varepsilon_0^2-8\varepsilon_0+5}}{8\varepsilon_0}\right)\\
&+\frac{\sqrt{5}}{5}\varepsilon\ln\left(\frac{5-4\varepsilon_0+\sqrt{5(4\varepsilon_0^2-8\varepsilon_0+5)}}{(11+5\sqrt{5})\varepsilon_0}\right)+16\varepsilon\ln\left(\frac{\sqrt{\varepsilon}+\sqrt{1-\varepsilon}}{\sqrt{\varepsilon}+\sqrt{\frac{\varepsilon}{\varepsilon_0}-\varepsilon}}\right)\\
&+8\varepsilon\frac{\sqrt{\varepsilon_0(1-\varepsilon_0)}+1-2\varepsilon_0}{\varepsilon_0+\sqrt{\varepsilon_0(1-\varepsilon_0)}}-8\sqrt{\varepsilon}\frac{\sqrt{\varepsilon(1-\varepsilon)}+1-2\varepsilon}{\sqrt{\varepsilon}+\sqrt{1-\varepsilon}}\Bigg).
\end{align*}
In the previous expression, $\varepsilon_0\approx 0.22815$ is the unique solution of the equation $$\displaystyle \frac{1}{1+2\sqrt{\varepsilon(1-\varepsilon)}} = \frac{-1 +\sqrt{4\varepsilon^2 -8\varepsilon + 5}}{2(1-\varepsilon)}.$$

\begin{figure}[th!]
\centering
 \includegraphics[scale=0.5]{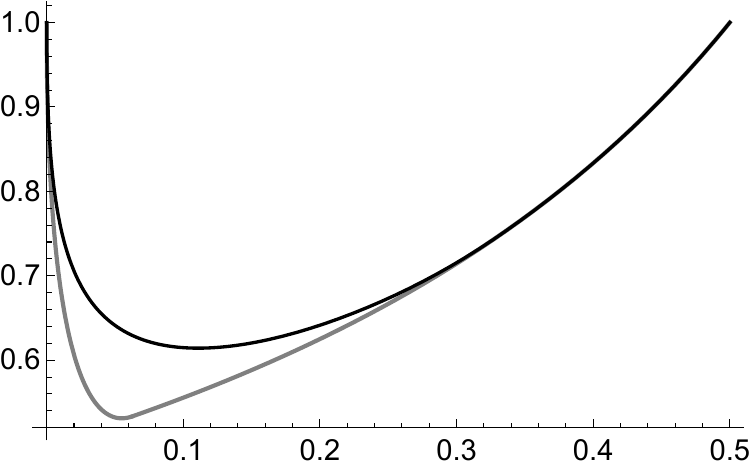}\\
  \caption{The top curve corresponds to the new approximation guarantee $G(\varepsilon)=F(\varepsilon)/(1-\varepsilon)$ with minimum $0.614247$; the gray curve below corresponds to the guarantee $H(\varepsilon)$ given  in \cite{Luca} with minimum 0.531: $H(\varepsilon)=(1-4\sqrt{\varepsilon}+8\varepsilon)/(1-\varepsilon)$, for $\varepsilon\leq 1/16$ and $H(\varepsilon)=  1/(2(1-\varepsilon))$, for $\varepsilon \geq 1/16$.} \label{plot2}
\end{figure}

\section{Conclusions} In this paper we have improved the analysis of Trevisan's spectral Max-Cut algorithm, showing that its approximation ratio is at least 0.614247. Furthermore, we have extended its applicability to the more general Maximum Colored Cut problem. 

We leave as an open problem to adapt the spectral algorithm and its  analysis to other problems that can be formulated as the maximization of a quadratic form over $\{-1,1\}$-vectors, such as Max-SAT, Max-$k$-SAT, Max-CUTGAIN \cite{Luca} and other constraint satisfaction problems. We believe that in many situations a spectral based algorithm can be useful to obtain good approximation algorithms without relying on the full power of semidefinite programming.

\end{document}